\documentclass[10pt,a4paper]{article}
\usepackage{amsthm}
\usepackage{amsmath}
\usepackage{amssymb}
\usepackage{amsfonts}
\usepackage{amstext}
\usepackage{amsbsy}
\usepackage[mathscr]{eucal}
\usepackage{graphicx}
\usepackage{color}
\usepackage{hyperref}

\usepackage{lettrine}
\usepackage{braket}

\numberwithin{equation}{subsection}
\makeindex
\title{\Large\textbf{Singularity Avoidance of Charged Black Holes in Loop Quantum Gravity}}
\author{\textit{ Mojtaba Taslimi Tehrani $^\dagger$  and  Hoshang Heydari}\\
\small\textit{Department of Physics, Stockholm University, 10691 Stockholm, Sweden}\\
\\\small\textit{$^\dagger$ Email: mota6264@student.su.se}\\}
\date{}
\begin{document}
\maketitle \thispagestyle{empty}
\begin{abstract}
Based on spherically symmetric reduction of loop quantum gravity, quantization of the portion interior to the horizon of a Reissner-Nordstr\"{o}m black hole is studied. Classical phase space variables of all regions of such a black hole are calculated for the physical case $M^2> Q^2$. This calculation suggests a candidate for a classically unbounded function of which all divergent components of the curvature scalar are composed. The corresponding quantum operator is constructed and is shown explicitly to possess a bounded operator. Comparison of the obtained result with the one for the Swcharzschild case shows that the upper bound of the curvature operator of a charged black hole reduces to that of Schwarzschild at the limit $Q \rightarrow 0$. This local avoidance of singularity together with non-singular evolution equation indicates the role quantum geometry can play in treating classical singularity of such black holes.
\end{abstract}
\maketitle

\section{Introduction}
While describes large scale structures of the universe remarkably precise, Einstein's theory of general relativity (GR) fails to give a consistent picture at certain space-time points, called singularity, where physical observables of the theory, such as energy density, become divergent. It is widely believed that the true cure of such physically meaningless states of gravity predicted by GR, lies out of the domain of validity of the theory; it must be resolved in a regime in which both gravitational and quantum mechanical effects are dominant.

Various attempts to construct a consistent theory describing such a regime has been made among them loop quantum gravity (LQG) \cite{Rovelli,Thiemann} presents a clear picture of quantum space-time at Planck scales in the kinematical level. The program rests on canonical quantization of GR expressed as an $SU(2)$ gauge theory on the 3 dimensional manifold $\Sigma$ of arbitrary topology, where a topology of space-time $M$ of the form $M\cong \mathbb{R} \times \Sigma $ is assumed, in a background independent manner. In such formulation of GR, the gravitational field is described by a pair of conjugate variables $(A, E)$, where $A^i_a(x)$\footnote{We are following Rovelli[1]'s notation in which indices $\mu, \nu ,...$ denote space-time, $i,j,...$ 3-space and $a,b,...$ $\mathbb{R}^3 $indices.} is an $SU(2)$ connection and $E^a_i(x)$ is the densitised triad vector field, conjugate to A:

 \begin{equation}
\{A^i_a(x) , E^b_j(x') \}= 8 \pi \gamma G \delta^i_j \delta^b_a \delta(x-x'),
 \end{equation}
with $G$ the gravitational constant and $\gamma$ the Immirzi parameter.

The conjugate pair are constraint to satisfy the system
\begin{eqnarray} \nonumber \label{cconstraint}
G_i= D_a E^a_i= 0,  \\ \nonumber
 H_b= E^a_i F_{ab} ^i= 0,\\ \nonumber
 H= \epsilon_{ijk} F_{ab}^{i} E^a_j E^b_k-2(1+\gamma^2) K^i_{[a}  K^j_{b]} E^a_i E^b_j =0,
 \end{eqnarray}
which are called Gauss, spatial diffeomorphism and Hamiltonian constraints respectively.
In fact, the task of finding a metric satisfying the Einstein's equations, describing configuration of a gravitating system, is now replaced by finding a conjugate pair $(A, E)$ satisfying the constraint system \ref{cconstraint}.

On quantization, one smears the basic fields $(A, E)$ to holonomies of $A^a_i$ along a curve $\gamma$, defined by $h_{\gamma} [A] =P \left( exp\int_\gamma A \right)$,  and fluxes of $E^a_i$ through the surface $S$, defined by $E_i(S)=\int_S d^2\sigma  n_a E^a_i$. They form the holonomy-flux algebra in which holonomies act by multiplication, and fluxes act by derivation.

Using a functional representation of quantum field theory and representing states as functionals of the cylindrical functions of holonomies, the kinematical Hilbert space of the theory is constructed. After imposing Gauss and diffeomorphism constraints as operators on such states, the true gauge and diffeomorphism invariant states of the theory turns out to be spin networks acted upon by holonomies and fluxes operators which form a unique representation. More precisely, a spin network is a triplet $(\Gamma, j_l, i_n)$ consisting of a graph $\Gamma$ with nodes in $\Sigma$, labeled by intertwiners $i_n$, and links connecting different nodes, labeled by $SU(2)$ representations $j_l$. The quantum geometrical picture suggested by LQG is manifest in quantization of geometrical observables, such as area and volume, as quantum operators acting on spin network states which result in discrete spectra and reflect the discrete nature of space-time.

While the kinematical construction of the theory is well understood, the dynamics is still poorly under control. This can be investigated either within the spin foam formalism, a path integral analogue for time evolution of spin networks, or by studying solutions to the Hamiltonian constraint, which requires criteria for choosing between different variants and orderings of the operator, and consequently building the dynamical Hilbert space. 

Despite difficulties in solving the dynamics of the full theory, different attempts have been made in studying simpler situations which would shed light on the path toward complete understanding of the theory. To study specific gravitational configurations characterized by specific macroscopic spatial symmetries, one should impose such symmetries suitably on the dynamical level of the quantum theory and find the true symmetric sector of the full theory. As for the Hydrogen atom in quantum mechanics, for instance, one would investigate the full quantum mechanical description of the system only by finding states satisfying the suitable dynamical equation, spherically symmetric Schr\"{o}dinger equation, which form basis of the dynamical Hilbert space. However, in LQG due to difficulties in solving Hamiltonian constraint, one employs the symmetry reduction methods \cite{Ashtekar-Bojowald}. In such models the desired symmetries are applied at classical level by choosing a symmetric pair $(A,E)$ satisfying constraints or it can be imposed at the kinematical level of the quantum theory. Quantization of the symmetric pair will then be done which, evidently, gives rise to a simpler procedure than in full theory. This is in fact the idea of mini superspace which has been used extensively in quantum cosmology\cite{Misner}.

For the homogeneous and isotropic cosmological models, loop quantum cosmology \cite{Bojowald3} has led to remarkable results such as replacing big bang singularity with a big bounce \cite{Bojowald4}, modification of Fredmann equation \cite{Taveras} and incorporation of an inflation mechanism \cite{Bojowald5}. However, because of the reduction of degrees of freedom to a finite number, in fact one, the field theoretical aspects of the gravitation are lost; since all points of space are equal, spin network just has one node. The same method has been applied to the Schwarzschild black hole, since the interior region of such a black hole is a homogeneous space of Kantowski-Sachs type, and its singularity is shown to be resolved \cite{Ashtekar2}.

Intuitively, singularity resolution occurs as a result of fundamental discreteness of space; while in a classical continuum, divergences emerge as distance goes to zero, there is no room for divergences in quantum level since there is no zero distance below the Planck length. Put it slightly different, in the same manner quantum mechanical model for the hydrogen atom prevents a classical electron from collapsing to the nucleus, and hence energy divergence, by putting a lower bound on the spectrum of energy, quantum geometrical model put an upper bound on the classically divergent curvature components and prevents black holes from forming a singularity.

In the spherically symmetric case, degrees of freedom of the theory cannot be reduced to a finite number. Instead, we are dealing with a midi-superspace having $U(1)$ as the gauge group. This makes calculations simpler than the full theory while still respects the field theoretical aspects of spherically symmetric quantum gravity. 

In this paper, we intend to study the quantum behavior of a Reissner-Nordstr\"{o}m black hole. To this end, we employ spherically symmetric quantum geometry methods \cite{Bojowald,Bojowald2} to quantize the interior portion of such a space-time within which the classical singularity occurs. In section 2, we introduce the states and operators used in spherically symmetric quantum geometry. Section 3 reviews the classical charged black hole metric, its horizons and singularity. And finally resolution of classical singularity will be discussed in section 4.

\section{Spherically Symmetric Quantum Gravity} \label{SSQG}

In this section, we outline the crucial points regarding the foundations and results of spherically symmetric quantum geometry developed by Bojowald \cite{Bojowald,Bojowald2}. The reader is referred to the original articles for many technical subtleties which are omitted inevitably.  

Viewing LQG as a method to quantize connections, one would be able to impose a symmetry through two avenues: (i) to pick, in the classical level, only those connections which are invariant under symmetry group action and consequently reduce the phase space, and (ii) to restrict the distributional states of the quantum theory, at the kinematical level, only to invariant connection.  
It is shown in \cite{Bojowald} that both methods just mentioned lead to the same result in the case of spherical symmetry. Therefore, we proceed by describing the former-that of quantizing the classically reduced phase space. 

\subsection{Classical Phase Space} \label{a}
The manifold $\Sigma$ having the symmetry group $S$ with an isotropy subgroup $F$, can be decomposed as $\Sigma \cong \Sigma / S \times S/ F$. The connection can generally be written as $A=A_{\Sigma / S}+A_{S / F}$ . Then $A_{\Sigma / S}$ can be considered as the connection of the reduced theory and its holonomies along curves in $\Sigma / S$ can be quantized.

For the spherically symmetric case, $\Sigma \cong \mathbb{R} \times S^2$,  and the symmetry group is $S=SU(2)$. This implies identifying $\Sigma / S$ with $\mathbb{R}$ and the gauge group of the reduced theory $F$ with $U(1)$. Therefore, reduced connections are $U(1)$ gauge fields on $\mathbb{R}$. Roughly speaking, spherical symmetry implies that our basic fields, in the spherical coordinate $(x, \theta, \phi)$, are independent of angular variables. Thus, the connection $A(\vec{x})$ is just a function of the radial coordinate; $A=A(x)$.

The connections and triads of the reduced spherically symmetric phase space have the general form:
\begin{equation}
A=A_x(x) \tau_3 dx + (A_1(x)\tau_1 + A_2(x) \tau_2) d\theta + ((A_1(x)\tau_2 - A_2(x) \tau_1) sin \theta + \tau_3 cos \theta) d\phi,
\end{equation}
\begin{equation} \label{Flux}
E=E^x(x) \tau_3 sin \theta \partial_x + (E^1(x)\tau_1 + E^2(x) \tau_2) sin \theta \partial_\theta + (E^1(x)\tau_2 - E^2(x) \tau_1) \partial_\phi,
\end{equation}
where $\tau_i=-\frac{i}{2} \sigma_i$ are the generators of $su(2)$ algebra. They define the symplectic structure:
\begin{equation}
\Omega = \frac{1}{2\gamma G} \int dx (dA_x \wedge dE^x + 2dA_1 \wedge dE^1 + 2 dA_2 \wedge dE^2).
\end{equation}

However, a suitable canonical transformation can be made resulting in canonical variables $(A_x(x), E^x(x))$, $(\gamma K_\phi(x), E^\phi(x))$ and $(\eta(x), P^\eta(x))$:

\begin{equation}
\Omega = \frac{1}{2\gamma G} \int dx (dA_x \wedge dE^x +  d(\gamma K_\phi) \wedge dE^\phi + 2 d\eta \wedge dP^\eta),
\end{equation}
with $K_\phi$ being the $\phi$ component of the extrinsic curvature.

The Gauss constraint, generating $U(1)$ gauge transformations, takes the form:

\begin{equation}
G[\lambda]= \int dx \lambda (E^{x\prime} +P^\eta)=0,
\end{equation}
 where prime denotes differentiation with respect to $x$.
 
Note that in terms of these variables, conjugate pair is not simply the connection-flux pair which suggests a different situation than the full theory.

The Hamiltonian constraint can be written as
\begin{equation} \label{H-constraint}
H[N]=-\frac{1}{2G} \int dx N(x) \frac{1}{\sqrt{|E^x|}} \left( (1- \Gamma_{\phi}^2+ K_{\phi} ^2)E^{\phi} + \frac{2}{\gamma} K_{\phi} E^x (A_x + \eta^{\prime})+ 2 E^x \Gamma_{\phi}^{\prime}\right).
\end{equation}
 
\subsection{Quantization}

Along the standard lines of constructing basic operators and states in the kinematical Hilbert space of LQG, one starts with  holonomies of the connections. Holonomies of $A_x $ along curves $\gamma$ in $\mathbb{R}$ are defined as $h^{(\gamma)} \equiv exp\left(\frac{i}{2}\int _\gamma A_x(x)\right)$ which are elements in $U(1)$. For $A_\phi$ point holonomies $ exp \left(i \mu A_\phi(x)\right)$ are used which belongs to the space of continuous almost periodic functions on the Bohr compactification of real line, and point holonomies of $\eta \in S^1$, have the form $ exp \left(i \eta(x) \right)$ which are elements of $U(1)$.

The kinematical Hilbert space of the present reduced theory is the space spanned by spin network state $T_{g, k, \mu}$
\begin{equation} \label{Spin-net}
T_{g, k, \mu}= \prod_{e \in g} exp\left(\frac{i}{2}k_e \int _e dx A_x(x)\right) \prod_{\nu \in V(g) } exp \left(i \mu_\nu \gamma K_\phi(\nu)\right) exp \left(i k_\nu \eta(\nu) \right).
\end{equation}
For a given graph $g$, these are cylindrical functions of holonomies along edges $e$ of $g$. Vertices $V(g)$ of such spin networks are labeled by irreducible $\bar{\mathbb{R}}_{Bohr}$ representations $\mu_\nu \in \mathbb{R}$ and irreducible $S^1$ representation $k_\nu \in \mathbb{Z}$, while edges are labeled by irreducible representations of $U(1)$. 

Holonomies act on spin network states by multiplication. Their corresponding momenta, on the other hand, act by differentiation
\begin{equation} \label{teraid x}
\hat{E}^x (x) T_{g, k, \mu} = \gamma \frac{\ell^2_p}{2} (k_{e^+(x)}+k_{e^-(x)})T_{g, k, \mu}, 
\end{equation}
\begin{equation}
\int dx \hat{E}^\phi (x) T_{g, k, \mu} = \gamma \ell^2_p \sum_v \mu_\nu T_{g, k, \mu} ,
\end{equation}
\begin{equation}
\int dx \hat{P}^\eta (x) T_{g, k, \mu} =2 \gamma \ell^2_p \sum_v k_\nu T_{g, k, \mu} .
\end{equation}

The volume operator can be expressed as $\hat{V}=4\pi \int dx |\hat{E}^\phi (x)| \sqrt{|\hat{E}^x (x)|}$ which is diagonal in spin network representation
\begin{equation} \nonumber
\hat{V} T_{g, k, \mu} = V_{k,m} T_{g, k, \mu};
\end{equation}
\begin{equation} 
V_{k,m}= 4\pi \gamma^{3/2} \ell_p^3 \sum_\nu |\mu_\nu|\sqrt{\frac{1}{2} |k_{e^+(x)}+k_{e^-(x)}|}.
\end{equation}

Implementing the Gauss constraint as an operator on spin networks to select the gauge invariant states, leads to a restriction on labels

 \begin{equation}
\hat{G}[\lambda] T_{g, k, \mu} = \gamma \ell_p^2 \sum_\nu \lambda(\nu) (k_{e^+(x)}-k_{e^-(x)}+2k_\nu)T_{g, k, \mu}
\end{equation}

 \begin{equation} \label{Gauss constraint}
\hat{G}[\lambda] T_{g, k, \mu} = 0 \Longrightarrow k_\nu = -\frac{1}{2}(k_{e^+(x)}-k_{e^-(x)}).
\end{equation}
Imposing this on \ref{Spin-net} results in the gauge invariant states
\begin{equation}
T_{g, k, \mu}= \prod_{e \in g} exp\left(\frac{i}{2}k_e \int _e dx ( A_x(x)+\eta^{\prime})\right) \prod_{\nu \in V(g) } exp \left(i \mu_\nu \gamma K_\phi (\nu)\right). 
\end{equation}

\section{Classical Reissner-Nordstr\"{o}m Black Hole}

The Reissner-Nordstr\"{o}m metric is a spherically symmetric solution to Einstein-Maxwell equations describing the space-time of a source with mass $M$ and charge $Q$ reads (in coordinate system $(x, \theta, \phi)$):
\begin{equation}\label{line element}
ds^2=-\left(1-\frac{2M}{x} + \frac{Q^2}{x^2}\right)dt^2+\left(1-\frac{2M}{x}+\frac{Q^2}{x^2}\right)^{-1}dx^2+x^2 d\Omega ^2.
\end{equation}
Horizons appear where $g_{00}=0$:
\begin{equation}
x^2-2Mx+Q^2=0 \longrightarrow x_{\pm}=M \pm \sqrt{M^2-Q^2}
\end{equation}
For the physically interesting case of $M^2>x^2$, event horizons partition space-time into 3 regions: I $(x>x_+)$, II $(x_- <x<x_+)$, and III $(0<x<x_-)$. By inspecting the sign of $g_{00}$, one observes that in region II, $x$ and $t$ interchange their roles and becomes time-like and space-like respectively. This defines a Kantowski-Sachs type homogeneous space. Region I and III, on the other hand, carry spherical symmetry.

\subsection{Classical Phase Space Variables}

\subsubsection{Region II}\label{II}

In region II, the metric of space-time takes the form
\begin{equation}
ds^2=-\left(\frac{2m}{t}-\frac{Q^2}{t^2}-1\right)^{-1}dt^2 + \left(\frac{2m}{t}-\frac{Q^2}{t^2}-1\right)dx^2+ t^2 d\Omega ^2.
\end{equation}
According to definition of tetrad (frame) fields,
\begin{equation} \label{region2 metric}
g_{\mu \nu}= \eta_{IJ} e^I_{\mu} e^J_{\nu},
\end{equation}
they can be determined only up to a Lorentz transformation. This leaves us with an $SO(3,1)$ freedom in choosing tetrad. In fact, given the metric \ref{region2 metric} we are free to choose their sign and Minkowski indices, which can be viewed as sort of a labeling 4 tetrad fields. However, in order to serve as the fundamental fields for constructing the conjugate pair $(A,E)$, a particular labeling must be chosen which will be clear below. 

The suitable choice for labeling 4 orthogonal frame fields turns out to be:
\begin{eqnarray}\nonumber \label{10}
e^0=\pm\left(\frac{2m}{t}-\frac{Q^2}{t^2}-1\right)^{-1/2}dt \textrm{ ; } e^1=\pm t sin\theta d\phi \\
e^2=\pm t d\theta \textrm{ ; } e^3=\pm \left(\frac{2m}{t}-\frac{Q^2}{t^2}-1\right)^{1/2}dx,
\end{eqnarray}
which leads to compatible spin connection components:
\begin{equation}
\omega^{30}=-\omega^{03}= \left(\frac{Q^2}{t^3}-\frac{m}{t^2}\right) dx \textrm{ ; } \omega^{20}= -\omega^{02}= \left(\frac{2m}{t}-\frac{Q^2}{t^2}-1\right)^{1/2} d\theta
\end{equation}
\begin{equation}
\omega^{10}=-\omega^{01}= \left(\frac{2m}{t}-\frac{Q^2}{t^2}-1\right)^{1/2}sin\theta d\phi \textrm{ ; }\omega^{12}=-\omega^{21}=cos\theta d\phi .
\end{equation}
The $A$ field can be constructed using spin connections:
\begin{equation}
A^3= \mp\gamma \left(\frac{Q^2}{t^3}-\frac{m}{t^2}\right) dx,
\end{equation}
\begin{equation}
A^2 = \pm \gamma  \left(\frac{2m}{t}-\frac{Q^2}{t^2}-1\right)^{1/2}d \theta,
\end{equation}
\begin{equation}
A^1 = \pm \gamma \left(\frac{2m}{t}-\frac{Q^2}{t^2}-1\right)^{1/2}sin\theta d\phi,
\end{equation}
\begin{equation}
A^3= \pm cos \theta d\phi.
\end{equation}

To construct the $E$ field on $\Sigma_{in}$ we choose a gauge in which $e^0_{\mu}=n_{\mu}$, the normal vector field to the spatial slice. This way we are in fact breaking the $SO(3,1)$ symmetry into $SO(3)$ on a hypersurface with topology $\Sigma= \mathbb{R} \times S^2$. The 3 triad fields become: 
\begin{equation} \label{3 teraid}
e^1=\pm t sin\theta d\phi \textrm{ ; } e^2=\pm t d\theta \textrm{ ; }e^3=\pm \left(\frac{2m}{t}-\frac{Q^2}{t^2}-1\right)^{1/2}dx,
\end{equation}
with determinant
\begin{equation}
\det e = t^2 sin \theta \left(\frac{2m}{t}-\frac{Q^2}{t^2}-1\right)^{1/2},
\end{equation}
and inverse triad 
\begin{equation}
e_1=\pm \frac{1}{t sin\theta} \partial_{\phi} \textrm{ ; } e_2=\pm \frac{1}{t} \partial_{\theta} \textrm{ ; } e_3=\pm \left(\frac{2m}{t}-\frac{Q^2}{t^2}-1\right)^{-1/2}\partial_x,
\end{equation}
The $E$ fields become:
\begin{equation}\nonumber
E_1 = \pm t \left(\frac{2m}{t}-\frac{Q^2}{t^2}-1\right)^{1/2} \partial_\phi, 
\end{equation}
\begin{equation}\nonumber
 E_2 =\pm t \left(\frac{2m}{t}-\frac{Q^2}{t^2}-1\right)^{1/2} sin\theta \partial_\theta,
\end{equation}
\begin{equation} \label{Ex}
E_3 =\pm t^2 sin\theta \partial_x .
\end{equation}

The 3 triad fields \ref{3 teraid} define their compatible spin connection, $\Gamma^{ij} \wedge e^j+de^i=0$
\begin{equation}
\Gamma ^{12}=-\Gamma ^{21}=cos \theta d \phi
\end{equation}
\begin{equation}
\Gamma^{3}=\frac{1}{2}\left(\epsilon^{312} \Gamma ^{12}+ \epsilon^{321} \Gamma ^{21}\right)=cos \theta d \phi.
\end{equation}

Extrinsic curvature is related to $A$ via $\gamma K=A-\Gamma$
\begin{equation}
K^3_r= \frac{1}{\gamma}A^3_r= \mp \left(\frac{Q^2}{t^3}-\frac{m}{t^2}\right) dx
\end{equation}

\begin{equation}
 K^2_\theta=\frac{1}{\gamma}A^2_ \theta= \pm \left(\frac{2m}{t}-\frac{Q^2}{t^2}-1\right)^{1/2}
\end{equation}

\begin{equation}
 K^1_\phi=\frac{1}{\gamma}A^1_ \phi= \pm \left(\frac{2m}{t}-\frac{Q^2}{t^2}-1\right)^{1/2}sin\theta
\end{equation}
Note that had we chosen other Minkowski indices for tetrad \ref{10} we would not have obtained the conjugate pair$(A,E)$ with correct indices satisfying $\{A^i_a(x) , E^b_j(x') \}= \delta^i_j \delta^b_a \delta(x-x')$.

The phase space variables are determined up to a sign freedom. By demanding $E$ and $A$ to satisfy the diffeomorphism, Gauss and Hamiltonian constraints, their signs can be fixed relative to each other.
All components of diffeomorphism and Gauss constraints are zero except:
\begin{equation}
H_\theta=\gamma t \left(\frac{2m}{t}-\frac{Q^2}{t^2}-1\right) cos \theta \left\{sgn(A^1_\phi)+sgn(A^2_\theta A^3_\phi)\right\},
\end{equation}
\begin{equation}
G_2= t \left(\frac{2m}{t}-\frac{Q^2}{t^2}-1\right)^{1/2}\cos \theta \left\{sgn(E^\theta_2 )+ sgn( A ^3_\phi E^{\phi1})\right\},
\end{equation}
and Hamiltonian constraint gives:
\begin{equation}
C=  t\left(\frac{2m}{t}-\frac{Q^2}{t^2}-1\right) sin^2\theta \left\{sgn(E^\theta_2) + sgn(E_1^\phi )\right\}.
\end{equation}
For the above constraints to be zero we must have:
\begin{equation}
sgn(E^\theta_2) = - sgn(E_1^\phi ),
\end{equation}
\begin{equation}
sgn(A^3_\phi) = +1,
\end{equation}
\begin{equation}
sgn(A^1_\phi) = - sgn(A^2_\theta ).
\end{equation}
This leaves us with two alternatives corresponding to the residual gauge freedom $(b, p_b) \rightarrow (-b, -p_b)$.
\begin{equation}
\textrm{I} \left\{ 
\begin{array}{l l}
A_a ^i= c \tau_{3} dr + b\tau_{2} d\theta +( cos\theta \tau_3 - b sin\theta \tau_1 )d\phi\\
E_i ^a= p_c \tau_3 sin\theta \partial_{r} + p_b \tau_2 sin \theta \partial_{\theta}-p_b \tau_1 \partial_ {\phi},\\
\end{array} \right.
\end{equation}
\begin{equation}
\textrm{II } \left\{ 
\begin{array}{l l}
A_a ^i= c \tau_{3} dr - b\tau_{2} d\theta +( cos\theta \tau_3 +b sin\theta \tau_1 )d\phi\\
E_i ^a= p_c \tau_3 sin\theta \partial_{r} - p_b \tau_2 sin \theta \partial_{\theta}+ p_b \tau_1 \partial_ {\phi},\\
\end{array} \right.
\end{equation}
where,
\begin{equation}\nonumber \label{131}
b=\pm \gamma \left(\frac{2m}{t}-\frac{Q^2}{t^2}-1\right)^{1/2} \textrm{ ; } c=\mp \gamma \left(\frac{Q^2}{t^3}-\frac{m}{t^2}\right)
\end{equation}
\begin{equation}\label{132}
p_c = \pm t^2 \textrm{ ; } p_b =t \left(\frac{2m}{t}-\frac{Q^2}{t^2}-1\right)^{1/2} .
\end{equation}
The momentum $p_c=\pm t^2$ is a monotonic function and can be interpreted as an internal time parameter (as is interpreted in \cite{Ashtekar2} for the case of the Kantowski-Sachs minisuperspace of Schwarzshild black hole).

\subsubsection{Region I and III}
The analoguos calculations for regions I and III with line element \ref{line element} leads to the folowing phase space coordinates 
\begin{equation} \label{Ex1}
\textrm{I} \left\{ 
\begin{array}{l l}
\tilde{A}_a ^i= \tilde{c} \tau_{3} dr + \tilde{b}\tau_{2} d\theta +( cos\theta \tau_3 - \tilde{b} sin\theta \tau_1 )d\phi\\
\tilde{E}_i ^a= \tilde{p}_c \tau_3 sin\theta \partial_{r} +\tilde{p}_b \tau_2 sin \theta \partial_{\theta}-\tilde{p}_b \tau_1 \partial_ {\phi},\\
\end{array} \right.
\end{equation}
\begin{equation}  \label{Ex2}
\textrm{II } \left\{ 
\begin{array}{l l}
\tilde{A}_a ^i= \tilde{c} \tau_{3} dr - \tilde{b}\tau_{2} d\theta +( cos\theta \tau_3 +\tilde{b} sin\theta \tau_1 )d\phi\\
\tilde{E}_i ^a= \tilde{p}_c \tau_3 sin\theta \partial_{r} - \tilde{p}_b \tau_2 sin \theta \partial_{\theta}+ \tilde{p}_b \tau_1 \partial_ {\phi},\\
\end{array} \right.
\end{equation}
where,
\begin{equation}\nonumber
\tilde{b}=\pm \gamma \left(1-\frac{2m}{x}+\frac{Q^2}{x^2}\right)^{1/2} \textrm{ ; } \tilde{c}=\mp \gamma \left(\frac{m}{x^2}-\frac{Q^2}{x^3}\right)
\end{equation}
\begin{equation}\label{13}
\tilde{p}_c = \pm x^2 \textrm{ ; } \tilde{p}_b =x\left(1-\frac{2m}{x}+\frac{Q^2}{x^2}\right)^{1/2} .
\end{equation}
This defines variables introduces in section \ref{a} as
\begin{equation}
A_x=\tilde{c} \textrm{ , } E^x = \tilde{p}_c \textrm{ ; }
\end{equation}
\begin{equation}
\gamma K_\phi= \tilde{b} \textrm{ , } E^\phi=\tilde{p}_b \textrm{ ; }
\end{equation}
\begin{equation}
\eta=(2n+1) \pi \textrm{ , } P^\eta=0,
\end{equation}
which constitute a 4 dimensional phase space.

The singularity occurs at $x=0$, or equivalently $E^x=0$, where the scalar curvature become infinite and lies in the spherically symmetric region III.

\section{Local Singularity Resolution: Curvature Boundedness }

In classical general relativity, a singularity is characterized by two criteria: (i) locally: the scalar curvature, made out of Riemann tensor, at a point of space-time becomes divergent, and (ii) globally: Einstein's equations cannot be solved for all values of $t$. Below, we consider the local criteria in the quantum theory and make a comment on the global one in the conclusion section. 

A look at Kretschmann scalar curvature of the Reissner-Nordstr\"{o}m black hole \cite{Kretschmann}
\begin{equation} \label{RNcurvature}
 R^{\mu \nu \rho \sigma} R_{\mu \nu \rho \sigma}= \frac{48 M^2 x^2 - 96 M Q^2 x + 56 Q^2 }{x^{8}},
\end{equation}
reveals that the irremovable curvature singularity of a charged black hole occurs at $x=0$. It lies in the spherically symmetric region III of the Reissner-Nordstr\"{o}m space-time. Our goal is to show that the quantum operator corresponding to $\frac{1}{x}$, on the other hand,  exhibits a spectrum which is bounded above. To construct such an operator, we need the classical function to be expressed suitably in terms of well-defined operators in the reduced theory. $\frac{sgn(E^x)}{ \sqrt{|E^x(x)|}}=\frac{1}{x}$ is a candidate to serve as the desired classical function. However, one cannot naively replace the inverse squired of triad with its operator analogue since, as can be seen in \ref{teraid x}, it has zero eigenvalue. Nevertheless, being expressed as a Poisson bracket of functions having well-defined operators, its corresponding quantum operator can be realized by replacing Poisson bracket with ($i \hbar$ times) commutator. 
This can be done by the simple relation: 
\begin{equation}
\mathcal{R} \equiv \frac{sgn(E^x)}{ \sqrt{|E^x(x)|}} = \frac{1}{2 \pi \gamma G}  \left\{A_x(x),  \sqrt{|E^x(x)|}\right\}.
\end{equation}

Since $\hat{E}^x(x)$ is diagonal on spin network states, $\sqrt{|\hat{E}^x(x)|}$ can be defined as a diagonal operator whose eigenvalues are square roots of (absolute value of) those of  $\hat{E}^x(x)$.

Along the inhomogeneous direction $x$, we can expand holonomy $h_x(x)$ as:
\begin{equation}
h_x=exp\left(i\int _e dx A_x(x)\right) \approx  1 + i\epsilon A_x(\nu), 
\end{equation} 
where $\epsilon=\nu_+ - \nu$ is the coordinate distance between two vertices $\nu$ and $\nu_+$ connected by the edge e. This enables us to, as is usual in LQG, express $\mathcal{R}$  to the order of $\epsilon^2$ as:
\begin{equation}
\mathcal{R} = \frac{1}{2 \pi \gamma G}  tr \left( \tau_3 h_x\{h_x^{-1}, \sqrt{|E^x(x)|}\}\right),
\end{equation}
and consequently its corresponding operator as:
\begin{equation}
\hat{\mathcal{R}} = \frac{1}{2 \pi \gamma \ell_{Pl}^2}  tr \left( \tau_3 \hat{h}_x \left[ \hat{h}_x^{-1}, \sqrt{|\hat{E}^x(x)|} \right]\right).
\end{equation}

To calculate the above expression, we write holonomy as:
\begin{equation}
h_x=exp\left(i\int _e dx A_x(x)\right)=cos(\frac{1}{2} \int A_x) + 2 \tau_3  sin(\frac{1}{2} \int A_x),
\end{equation}
and therefore,
\begin{eqnarray} \nonumber
\hat{h}_x\left[ \hat{h}_x^{-1}, \sqrt{|\hat{E}^x(x)|} \right] &=& \sqrt{|\hat{E}^x(x)|} - cos(\frac{1}{2} \int A_x)\sqrt{|\hat{E}^x(x)|}cos(\frac{1}{2} \int A_x)\\ \nonumber
&& -  sin(\frac{1}{2} \int A_x)\sqrt{|\hat{E}^x(x)|}sin(\frac{1}{2} \int A_x) \\ \nonumber
&& +2 \tau_3 \left( cos(\frac{1}{2} \int A_x)\sqrt{|\hat{E}^x(x)|}sin(\frac{1}{2} \int A_x)\right.\\ \nonumber
&&  \left. -  sin(\frac{1}{2} \int A_x)\sqrt{|\hat{E}^x(x)|}cos(\frac{1}{2} \int A_x) \right). \nonumber
\end{eqnarray}
This leads to 
\begin{equation}
\hat{\mathcal{R}} =\frac{1}{2 \pi \gamma \ell_{Pl}^2} \left( cos(\frac{1}{2} \int A_x)\sqrt{|\hat{E}^x(x)|} sin(\frac{1}{2} \int A_x) -  sin(\frac{1}{2} \int A_x)\sqrt{|\hat{E}^x(x)|} cos(\frac{1}{2} \int A_x)  \right).
\end{equation}

To investigate its action on spin network states, one can use identities expressing $sin x $ and $cos x$ in terms of exponentials and write:
\begin{eqnarray} \nonumber
\hat{\mathcal{R}} T_{g, k, \mu}&=&  \frac{1}{2 \pi \gamma \ell_{Pl}^2} \left( (e^{\frac{i}{2} \int A_x}+e^{-\frac{i}{2} \int A_x})\sqrt{|\hat{E}^x(x)|}(e^{\frac{i}{2} \int A_x}-e^{-\frac{i}{2} \int A_x}) \right. \\ \nonumber
&& \left. - (e^{\frac{i}{2} \int A_x}-e^{-\frac{i}{2} \int A_x})\sqrt{|\hat{E}^x(x)|}(e^{\frac{i}{2} \int A_x}+e^{-\frac{i}{2} \int A_x}) \right)\\ \nonumber
&& \times \prod_{e, \nu} e^{\left(\frac{ik_e }{2}\int _e ( A_x+\eta^{\prime})\right)} e^{ \left(i \mu_\nu \gamma K_\phi \right)}, \nonumber
\end{eqnarray}
which gives rise to the spectrum

\begin{eqnarray} \nonumber
\hat{\mathcal{R}} T_{g, k, \mu} = \frac{1}{2  \pi \sqrt{\gamma} \ell_{Pl}} \left(\sqrt{\frac{1}{2} |k_{e^+(x)}+k_{e^-(x)}+2|}-\sqrt{\frac{1}{2} |k_{e^+(x)}+k_{e^-(x)}-2|} \right)T_{g, k, \mu}.\\
&&
\end{eqnarray}
\begin{figure} \label{RNss}
\centering
\includegraphics[width=0.6\textwidth]{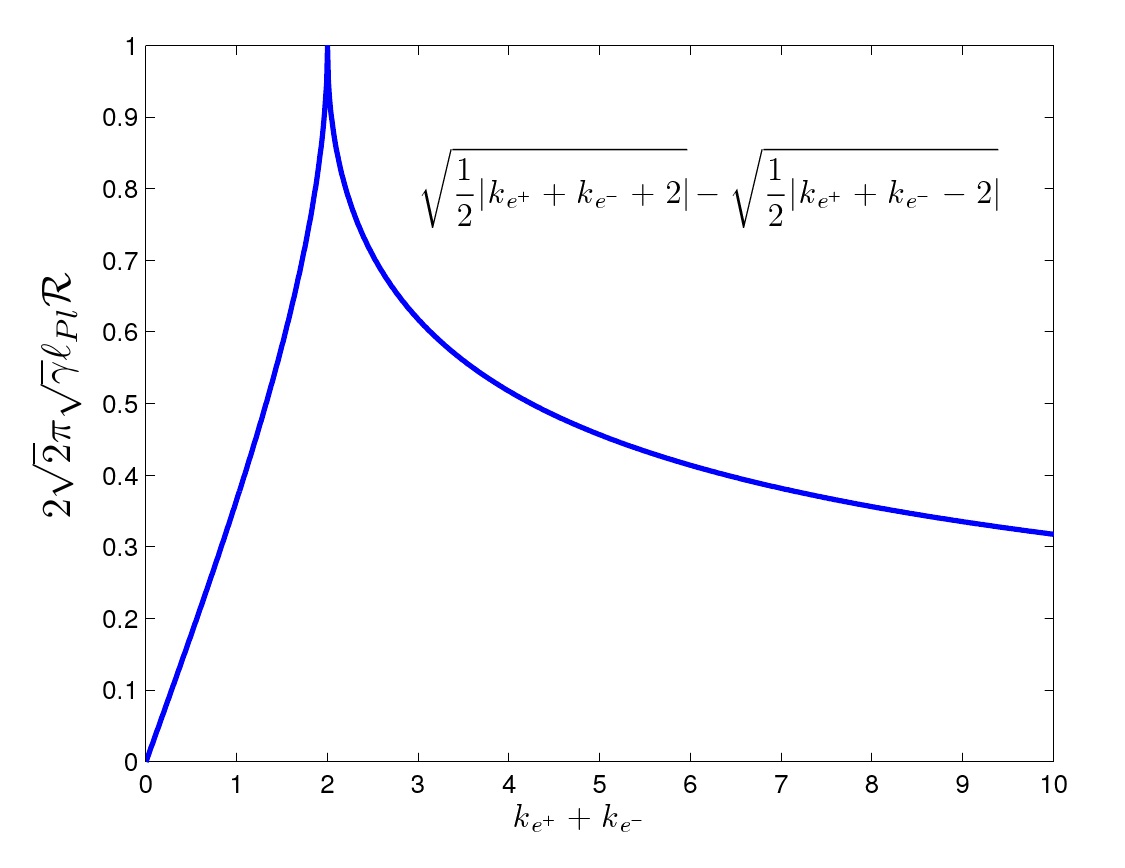}
\caption{The upper bounded spectrum of $\frac{1}{x}$ for a charged black hole}
\end{figure}
 
Such an operator exposes two features: (i) it is diagonal on spin network states and therefore commutes with all operators in kinematical Hilbert space, and (ii) it has a bounded spectrum (see figure 1) with maximum value of $ \frac{1}{\sqrt{2} \pi \sqrt{\gamma} \ell_{Pl}}$. Thus, the scalar curvature \ref{RNcurvature}, which is classically divergent, at quantum level has a maximum value of:
\begin{eqnarray} \nonumber \label{RNcurv}
\left( R^{\mu \nu \rho \sigma} R_{\mu \nu \rho \sigma}\right)_{max}&=& \left(\frac{48 M^2 x^2 - 96 M Q^2 x + 56 Q^2 }{x^{8}}\right)_{max}\\ 
&=& \frac{1}{\gamma^3 \pi^6 \ell^{6}_{Pl}} \left( 6 M^2  - \frac{96  M Q^2}{\sqrt{2} \pi \sqrt{\gamma} \ell_{Pl}}+ \frac{28 Q^2}{ \pi^2 \gamma \ell^2_{Pl}}\right).
\end{eqnarray}. 
This reflects the fact that the divergence of $\frac{1}{x}$ is kinematically well-behaved at quantum level.


\section{Comparison with Schwarzschild Case}

The Kretschmann scalar curvature of the Schwarzschild black hole, $ R^{\mu \nu \rho \sigma} R_{\mu \nu \rho \sigma}= \frac{48 M^2}{t^6}$, indicates that in this case the irremovable singularity of space-time lies at $t=0$ as well. However, the region within the event horizon of such a space-time is of a Kantowski-Sachs type and methods introduced in section \ref{SSQG} are not applicable to it. Loop quantization of such space-times, which are of interest in cosmology as well, is discussed in \cite{Ashtekar2}. However, in \cite{Ashtekar2} the issue of resolution of classical singularity in quantum regime is addressed by illustrating the stronger condition of non-singular behavior of the quantum dynamical equation, and the boundedness of scalar curvature operator is not exhibited directly. Below, we sketch the construction of  such an operator in a manner similar to the calculations of the previous sections.

The classical phase space variables would be the same calculated in subsection \ref{II} in the limit $Q=0$. Therefore the set $\{c, p_c\}$ used in this section is meant to be understood as the $Q=0$ limit of \ref{131} and \ref{132}.

Consider the following quantity on the classical phase space
\begin{equation}
 \mathcal{R}\equiv \frac{1}{2 \pi \gamma G} \left\{c,  \sqrt{|p_c|}\right\} = \frac{sgn(p_c)}{ \sqrt{|p_c|}}=\frac{1}{t}.
\end{equation} 

Following the methods presented in \cite{Ashtekar2}, expand the holonomy along $x$ direction of $\Sigma=\mathbb{R} \times S^2$ with oriented length $\tau$ as
 \begin{equation} 
h_{x}^{(\tau)}= 1 + \epsilon \int_0^\tau dx c \tau_3 +\mathcal{O}(\epsilon^2),
\end{equation}
and write $\mathcal{R}$ as
\begin{equation}
 \mathcal{R}=  \frac{1}{2 \pi \gamma G}  tr  \left( \tau_3 h_x^{(\tau)} \left\{h_x^{(\tau) -1},  \sqrt{|p_c|}\right\} \right). 
\end{equation} 

Now, quantization would be straightforward:

\begin{eqnarray} \nonumber \label{Rhat}
 \hat{\mathcal{R}} &=&  \frac{1}{2 \pi \gamma \ell_{Pl}^2}  tr  \left( \tau_3 \hat{h}^{(\tau) }_x \left[\hat{h} _x^{(\tau) -1},  \sqrt{|\hat{p}_c|}\right ] \right) \\ \nonumber
&=& \frac{1}{2 \pi \gamma \ell_{Pl}^2} \left( cos(\frac{\tau c}{2} )\sqrt{|\hat{p}_c|}sin(\frac{\tau c}{2}  ) -  sin(\frac{\tau c}{2}  )\sqrt{|\hat{p}_c|}cos(\frac{\tau c}{2}  ) \right).\\
\end{eqnarray} 

Its action on $\ket{\tau, \mu}$, which are the simplified version of the spin network states in this reduced model (with $\mu$ being the oriented length along the equator of $S^2$), then becomes:

\begin{equation}\label{Rmu} 
 \hat{\mathcal{R}} \ket{\tau, \mu} =  \frac{1}{2\pi \sqrt{\gamma} \ell_{Pl}}  (\sqrt{|\tau+1|}- \sqrt{|\tau-1|}) \ket{\tau, \mu}.
\end{equation}

Such an operator has a bounded spectrum with maximum value of $ \frac{1}{\sqrt{2}\pi \sqrt{\gamma} \ell_{Pl}}$. This means that the scalar curvature, which is classically divergent, at quantum level has a maximum value of:
\begin{equation}  \label{Smax}
\left( R^{\mu \nu \rho \sigma} R_{\mu \nu \rho \sigma}\right)_{max}= \left(\frac{48 M^2}{t^6}\right)_{max}= \frac{6 M^2}{\gamma^3 \pi^6 \ell^6_{Pl}}.
\end{equation} 

Note that in the limit $Q \rightarrow 0$, the maximum value of the curvature of a Reissner-Nordstr\"{o}m black hole, \ref{RNcurv}, reduces to that of Schwarzschild black hole \ref{Smax}.


\section{Conclusion}
The remarkable result in quantum geometry is the discreteness of space manifested in discrete spectrum of geometrical observables such as area and volume operators. This enables us to express other classically ill-defined functions in terms of basic quantized fields, and to study their quantum mechanical behavior. One of such functions is the scalar curvature of a  solution of Einstein's equations. In cosmological models, divergence of the scalar curvature is due to unboundedness of inverse scale factor at initial singularity, while in black holes this happens at the center of black hole where the function $\frac{1}{x}$, constituting all curvature components, becomes arbitrary large. In the above sections, we have constructed the quantum analogue of $\frac{1}{x}$ for the spherically symmetric region interior to the horizons of a Reissner-Nordstr\"{o}m black hole, and have shown that it is an operator possessing a bounded spectrum. However, this is just the local criterion for absence of singularity; the global criterion requires the quantum Einstein's equation representing the dynamics to be non-singular as well. To study the dynamics, we must encode time evolution coded in Hamiltonian constraint. To this end, one must assign a time parameter to the theory. Such a quest is usually realized via coupling a scalar field to the theory. However, in the case of spherical symmetric space of a charged black hole, we can consider the monotonic function $E^x(x)= \pm x^2$ as an ``internal time''. Since for the interior portion of a Reissner-Nordstr\"{o}m space (region II) the internal time is taken as $\pm t^2$, this strongly suggests, for the sake of continuity, to take the same function $\pm x^2(=\pm t^2)$ to serve as this purpose. In the quantum level, thus, the eigenvalues of $\hat{E}^x(x)$ play the role of discrete internal time. To see whether or not the time evolution of quantum space-time come to a halt at singularity $E^x(x)=0$, we must act by Hamiltonian constraint on spin network states and inspect whether or not the coefficients of its eigenvalues remain non-vanishing at such a point. The main task of realizing the Hamiltonian operator is done and extensively discussed in \cite{Bojowald2}, and as pointed out in \cite{Bojowald7}, among the various options to choose an ordering for Hamiltonian constraint, the symmetric one, $\hat{H}=\hat{H}_{\nu}^{\dagger}+\hat{H}_{\nu}$, turns out to be the one with non-vanishing coefficients and can serve as the criterion of non-singular evolution equation. Therefore, absence of singularity can also play the role of a criterion for selecting the right ordering for the theory.

\end{document}